\documentstyle[prl,aps,epsbox]{revtex}

\begin{document}
\twocolumn

\def\slashchar#1{\setbox0=\hbox{$#1$}           
   \dimen0=\wd0                                 
   \setbox1=\hbox{/} \dimen1=\wd1               
   \ifdim\dimen0>\dimen1                        
      \rlap{\hbox to \dimen0{\hfil/\hfil}}      
      #1                                        
   \else                                        
      \rlap{\hbox to \dimen1{\hfil$#1$\hfil}}   
      /                                         
   \fi}                                         %

\title{Lattice representation of vector and chiral gauge theories}
\author{Takanori Sugihara}
\address{
RIKEN BNL Research Center, 
Brookhaven National Laboratory, Upton, New York 11973, USA
}

\maketitle

\begin{abstract}
A lattice derivative is defined as 
a discrete Fourier transform of momentum on a finite lattice. 
Species doublers are removed with anti-periodic boundary conditions. 
U(1) chiral transformation is modified to reproduce chiral anomaly. 
Chiral gauge theories can be constructed on the lattice 
using a single Weyl fermion as a building block. 
See Refs. \cite{Sugihara:2003ga,Sugihara:2003mh}. 
\end{abstract}

\hspace{-0.2cm}

Our doubler-free lattice derivative is defined as follows. 
\begin{eqnarray}
 \nabla_n &=& \frac{1}{N} \sum_{l=-N/2+1}^{N/2}
 ip_l e^{i2\pi \tilde{l}n/N}
\nonumber
\\
 &=& \frac{\pi}{N^2}
 \left[
 (N+1)
 \frac{\cos\left(\displaystyle\frac{Ns}{2}\right)}
      {\sin\left(\displaystyle\frac{s}{2}\right)}
  -\frac{\sin\left(\displaystyle\frac{(N+1)s}{2}\right)}
       {\sin^2\left(\displaystyle\frac{s}{2}\right)}
 \right],
\label{der}
\end{eqnarray}
where $p_l\equiv 2\pi\tilde{l}/N$, $\tilde{l}\equiv l-1/2$, 
and $s\equiv 2\pi n/N$. 
The derivative has antiperiodicity, $\nabla_{n+N}=-\nabla_n$. 
In the continuum limit $a\to 0$ after $N\to\infty$, 
we obtain the first order 
derivative of the continuum theory.
\begin{equation}
 \lim_{a\to 0}\frac{1}{a}\nabla_n 
 = a \frac{\partial}{\partial x} \delta(x). 
 \label{nabla3}
\end{equation}
The derivative has an appropriate continuum limit. 
The lattice covariant derivative
\begin{equation}
 (D_\mu)_{m,n} \equiv \frac{1}{a}
  \nabla_{m_\mu-n_\mu} U_{m,n}(\mu)
  \prod_{\nu=1 (\nu\ne\mu)}^4 \delta_{m_\nu,n_\nu}
\label{lcd}
\end{equation}
is diagonal with respect to the space-time indices $m$ and $n$ 
except for the $\mu$-th ones. 
The variable $U_{m,n}(\mu)$ is a product of all link variables 
that compose a line segment between the two sites $m$ and $n$ 
parallel to the $\mu$-th direction. 
The Dirac operator is a simple matrix easy to implement to a computer. 

To simulate chiral anomaly on the finite lattice, 
we introduce the modified chiral transformation with $\theta_n\ll 1$ 
\begin{eqnarray}
 \psi'_m &=&
 \sum_n \left(1+i\theta_m\hat{\gamma}_5\right)_{m,n}\psi_n,\quad
\\
 \bar{\psi}'_m &=&
 \sum_n \bar{\psi}_n\left(1+i\theta_m\hat{\gamma}_5\right)_{n,m},
\end{eqnarray}
where 
$(\hat{\gamma}_5)_{m,n} \equiv \gamma_5 \left(1-a G/2 \right)_{m,n}$. 
The operator $G$ is the Neuberger's solution \cite{neuberger} to the 
Ginsparg-Wilson relation $\gamma_5 G + G \gamma_5 = aG \gamma_5 G$ 
and has nothing to do with the Dirac operator (\ref{lcd}). 
Under global transformation with $\theta_n=\theta$, 
the classical action $S$ generate an explicit breaking term 
proportional to lattice spacing $a$. 
\begin{equation}
 \delta S = \frac{i}{2} \theta a^4 \sum_{m,n}
 \bar{\psi}_m \gamma_5 a(\slashchar{D}G-G\slashchar{D})_{m,n}\psi_n. 
\label{var}
\end{equation}
In the continuum limit, the breaking term vanishes 
and the global transformation is a symmetry of 
the classical action. 
However, the breaking term reproduces topological charge 
quantum mechanically and gives the index theorem 
for arbitrary lattice spacing. 
The explicit breaking term (\ref{var}) is necessary for 
taking care of zero mode of the anomalous Ward identity as shown below. 

The axial current divergence is defined as a variation 
of the classical action under the local transformation. 
\begin{eqnarray}
 (\partial_\mu \hat{J}^5_{\mu})_m&=&\sum_{n_1,n_2}\Big[
  \bar{\psi}_{n_1}(\hat{\gamma}_5)_{n_1,m}
  \slashchar{D}_{m,n_2}\psi_{n_2}
\\
  &&+\bar{\psi}_{n_2}
  \slashchar{D}_{n_2,m}(\hat{\gamma}_5)_{m,n_1}\psi_{n_1}
 \Big].
\end{eqnarray}
The fermion measure transforms as
\begin{equation}
 {\cal D}\psi' {\cal D}\bar{\psi}' =
 \exp\left(-2i\sum_n \theta_n {\cal A}_n \right)
 {\cal D}\psi {\cal D}\bar{\psi}. 
\end{equation}
where chiral anomaly 
${\cal A}_n\equiv {\rm tr} (\hat{\gamma}_5)_{n,n}$ 
is a gauge-invariant quantity. 
Then, we have the anomalous Ward identity 
\begin{equation}
 \langle (\partial_\mu \hat{J}^5_{\mu})_n \rangle
 =-\frac{2}{a^4} {\cal A}_n, 
\end{equation}
which holds also for the zero mode of the axial current divergence. 
As shown in Ref. \cite{luscher},  
the index theorem holds for arbitrary lattice spacing. 
\begin{equation}
 \sum_n {\cal A}_n={\rm Index}(G). 
\end{equation}

A single Weyl fermion can exist on the lattice 
without violating gauge symmetry 
because the fermion measure is invariant under 
gauge transformation. 

The advantage of the proposed method is the complete existence of 
the anomalous Ward identity and exact flavor chiral symmetry. 
Approximation of $\hat{\gamma}_5$ does not modify flavor chiral symmetry.

\end{document}